# A Brief Overview of the MAD Debugging Activities[1]

D. Kranzlmüller, Ch. Schaubschläger, J. Volkert
GUP Linz, Joh. Kepler University Linz
Altenbergerstr. 69, A-4040 Linz
kranzlmueller@gup.uni-linz.ac.at
http://www.gup.uni-linz.ac.at/

**Abstract**

Debugging parallel and distributed programs is a difficult activity due to the multiplicity of "sequential bugs", the existence of malign effects like race conditions and deadlocks, and the huge amounts of data that have to be processed. These problems are addressed by the Monitoring And Debugging environment MAD, which offers debugging functionality based on a graphical representation of a program's execution. The target applications of MAD are parallel programs applying the standard Message-Passing Interface MPI, which is used extensively in the high-performance computing domain. The highlights of MAD are interactive inspection mechanisms including visualization of distributed arrays, the possibility to graphically place breakpoints, a mechanism for monitor overhead removal, and the evaluation of racing messages occurring due to nondeterminism in the code.

## 1. Introduction

Parallel programs consist of several, concurrently executing and communicating tasks, which jointly solve a given problem. In message-passing programs, the communication and synchronization between these tasks is carried out with dedicated functions for message transmission. Consequently, the traditional software life-cycle has to be adapted to cope with concurrency and process interaction. This exhibits problems unknown in sequential computing, which must be addressed during software design, implementation, and testing and debugging.

The testing and debugging phase is responsible for detecting erroneous behavior and locating its original reason as established by the application's source code. This is achieved with methods like breakpointing and inspection, which are applied during re-executions of the target program to comprehend its behavior [10]. A parallel debugging tool must adapt these basic methods to the characteristics of the parallel program and may additionally supplement the error detection activities by facilities addressing problems established by parallelism.

Both goals are covered by the **M**onitoring **A**nd **D**ebugging environment **MAD**, which is a toolset for debugging parallel programs based on the message-passing paradigm. The principal idea of MAD is to display observed program interactions as a space-time diagram and apply this graphical representation as the main debugging abstraction. The advantage of this approach compared to other debugging environments is improved program understanding.

---

Some well-known parallel debuggers which are based on a textual representation of the program's source code are DDBG [1], P2D2 [5], and Totalview [3]. These debuggers may be inadequate for parallel program analysis due to the inherent complexity of concurrently executing and communicating processes [9]. Problems with the complexity and the amount of data are also described with a phenomenon called the maze-effect [2], which offers another justification for debugging with a graphical abstraction.

Based on the graphical representation, MAD offers a variety of debugging features, for example placement of breakpoints on multiple processes, inspection of variables, communication events, and composite data structures like distributed arrays. Furthermore, MAD includes functionality for critical errors occurring in context with nondeterministic program behavior. The consequences of race conditions can be evaluated with an integrated record&replay mechanism and a sophisticated event manipulation feature.

This paper represents a brief overview of the debugging features included in MAD. The next section briefly introduces the monitoring part, that observes a program's execution. Section 3 shows how the observed data is presented to the user. Afterwards, the features dedicated to nondeterminism analysis are described, before concluding with an outlook on future work in this project. In addition to this paper, we will present this overview or the usage of MAD, respectively, in a demonstration session. The contents of the demo are described in the Appendix of the paper. The contribution of this paper is on the one hand, the introduction of novel ideas like array inspection and automatic nondeterminism testing, and on the other hand, an overview of the connection between these features, that has not been described before.

## 2. Monitoring Parallel Programs

The initial step for program debugging is instrumentation, which is applied to insert monitoring functionality into the target program, so that its execution can be observed and information for the analysis step is retrieved. In the MAD environment, monitoring is performed by the **NO**ndeterministic **P**rogram **E**valuator **NOPE**, which is inserted into the target program through source code instrumentation.

With NOPE, debugging data is generated whenever an event of interest occurs. Two different kinds of events can be distinguished, those that are automatically generated, and those that are inserted by the user. For example, communication events as established by the message passing interface are automatically instrumented and observed by NOPE. Artificial events, like variable and message queue inspection, have to be inserted manually by the user. An example for an artificial type of event is array inspection, which is included in the program by placing the macro *monARRAYTRACE*(data,info).The first parameter of this macro is a pointer to the data array, while the second parameter is a structure that describes the contents of the array, like element type and size, process array size, and array distribution.

After the source code has been instrumented, the program is recompiled and linked with NOPE. During successive program executions trace files will be generated for post-mortem analysis, containing information about occurring events. These traces include event number, event type and timestamps for all types of events, message length for communication events, variable contents for inspection events, etc., and can be used for debugging as well as for performance evaluation [6].

The collected information about the program run must be used with care, as monitoring of programs introduces some problems. One problem is the probe effect [4], which means that the results of any measurement will always be imprecise since the tool used for the measurements will influence the observed target. In case of a software monitor like NOPE the mea-surements are affected by the additional code that is added to the target program.

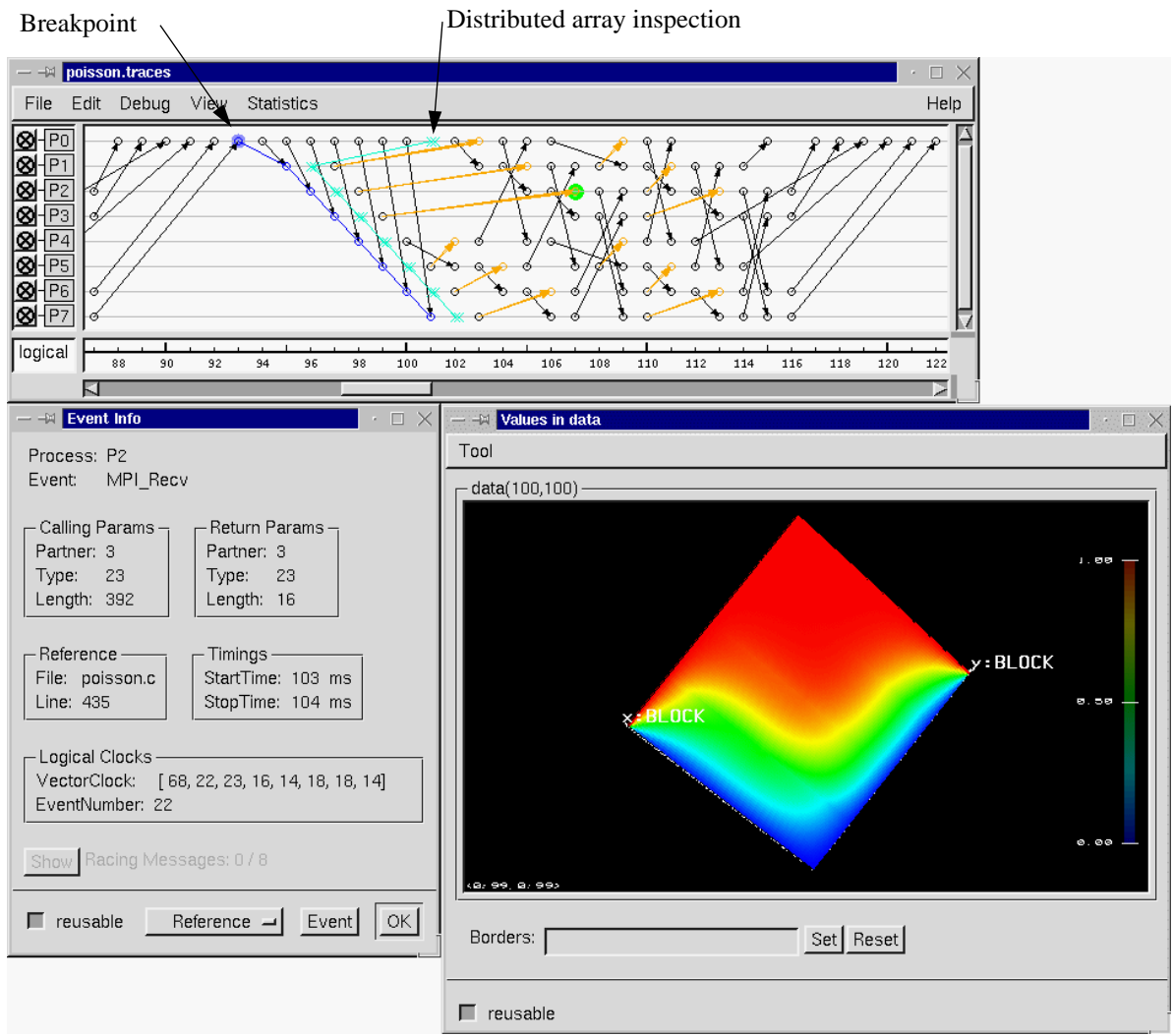

Figure 1. Event graph visualization for parallel program debugging

Another problem is established by the fact, that real-time clocks in a parallel computer are often not synchronized, and it is impossible to determine a global order of events. Possible results are timings, that indicate that a message was received before it was even transmitted. Both problems are addressed in MAD where a dedicated tool automatically performs post-mortem clock synchronization and removes monitor overhead from the traces [11].

## 3. Visualization of Parallel Programs

After the trace data has been collected it is presented to the user with **ATEMPT**, **A T**ool for **E**vent **M**ani**P**ula**T**ion, as a space-time diagram [7]. This higher level of abstraction provides the basis for further program analysis and debugging features.

Firstly, ATEMPT automatically detects and highlights simple communication errors like isolated events and communication events with different message length at sender and receiver. By selecting an event in the space-time diagram a window with additional information is opened. With an integrated source code reference the graphical object can be related to the original lines of code, thus, compensating for the higher level of abstraction.

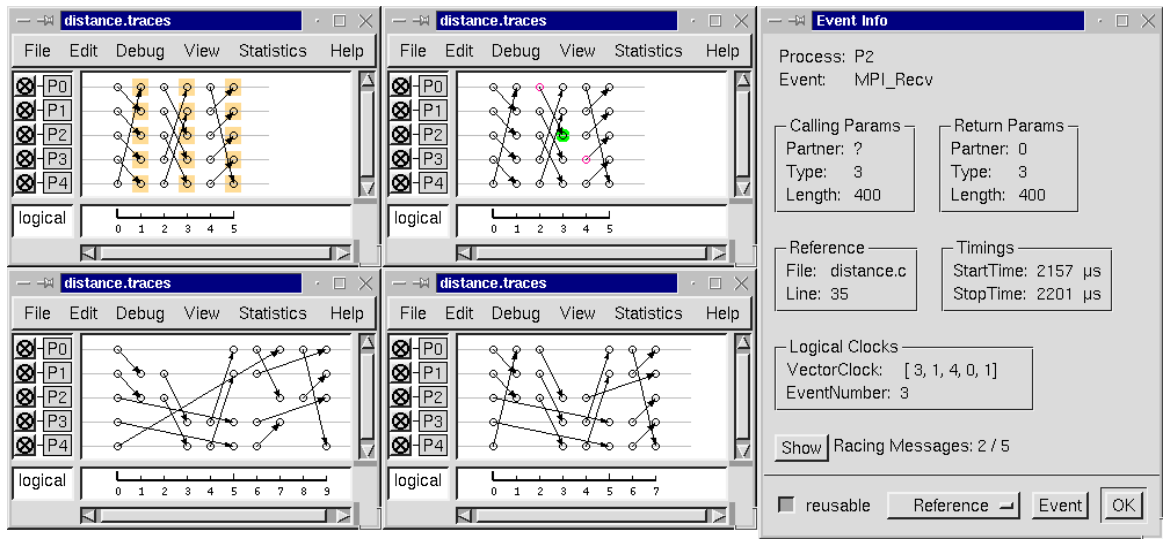

Figure 2. Nondeterminism analysis with MAD

Support for distributed array inspections as described above is included in the visual representation with a dedicated graphical object. Selecting this object opens a window with meta-information about the array. In addition, the contents of the array can be displayed as a heat diagram, and it is also possible to visualize the mapping of the data array onto the processes. With this visualization, computational errors in arrays can be detected immediately.

Another important feature of ATEMPT its capability to set breakpoints in the graphical display. Therefore, the user selects an appropriate event on one process, and ATEMPT automatically evaluates at which events on the other processes the program's execution must be halted. These breakpoints are selected as early as possible in order to provide the maximum freedom for the user. They come into effect during program re-executions, and allow users to attach other debuggers for traditional debugging activities.

Figure 1 contains examples for each of these features. The space-time diagram for an arbitrary parallel program contains highlighted erroneous events, a distributed array inspection, and a breakpoint across all processes. The bottom-left event info window shows the data associated with a selected event on process P2. The bottom-right window shows a heat-diagram for the selected distributed array.

## 4. Evaluation of Nondeterministic Behavior

Another purpose of MAD is analysis of nondeterministic parallel programs. A program is nondeterministic if it produces different results in successive program runs even if the same input data are provided [8]. In message passing programs the main source of nondeterminism are race conditions, which occur at receive events if two conditions are fulfilled: Firstly, the source of the received message is unspecified, so that a message from any source can be accepted. This is called a wild card receive and is accomplished in MPI by using the constant *MPI_ANY_SOURCE* as a specification for the message's origin. Secondly, there must be at least two so-called racing messages that can be accepted at that particular wild card receive. In such a case it is unpredictable, which one of the racing messages will be accepted during program execution. Thus, in successive program runs the order of received messages might be different, which possibly leads to different program behavior. This makes debugging an annoying task due to two problems: the irreproducibility effect, which means that equivalent re-execution of a nondeterministic program cannot be guaranteed, and the incompleteness problem, which means that often only a subset of all possible execution paths can be tested.

For that reason, the MAD environment provides an event manipulation feature and a record&replay mechanism. Firstly, all wild card receives in the space-time diagram are highlighted. Secondly, for a selected wild card receive, all corresponding racing messages are determined. Thirdly, with event manipulation the user may graphically change the observed program behavior by choosing one of the racing messages to arrive at the wild card receive. The idea of this strategy is to evaluate the question: What would have happened, if the order of events would have been different?

This questions is evaluated during a re-execution of the program under control of the record&replay mechanism. To see the effects of the exchange, the program has to be re-executed as follows:

- before the user-selected wild card receive event the program is replayed as observed during a previous program execution.
- at the selected wild card receive the specified message will be accepted as defined by event manipulation
- after the exchange the program is re-executed without any constraints, since it is impossible to make predictions about the future of the program after the alteration at the nondeterministic receive.

Examples for the nondeterminism analysis with MAD are shown in Figure 2. The top-left diagram displays a simple parallel program with all its wild card receives. In the top-middle diagram, one of the wild card receives has been selected and all racing messages are highlighted. An event info window for this receive event is shown in the right-most screenshot, which displays the wild card at the partner ("?") and the number of racing messages. The two bottom diagrams show two other possible executions of the same parallel program.

## 5. Conclusion and Future Work

The MAD environment is a debugging toolset for parallel programs based on the message-passing paradigm. In contrast to most other tools in this domain, MAD is centered around a graphical representation of the program flow, which is used for in-depth analysis activities like breakpointing, event inspection, and array visualization. In addition, the combination of record&replay mechanisms with event manipulation allow to evaluate the consequences of race conditions in parallel programs.

The biggest benefit of our approach seems to be a reduction of the analysis complexity, which mainly stems from the graphical representation and its application as the central debugging interface. Other advantages like race condition detection and monitor overhead removal have been proved useful for a variety of critical situations.

The future goals in this project can be divided into three distinct parts. One goal is the graphical representation of other complex data structures besides multi-dimensional arrays. Another issue is the integration of traditional debugging services into the debugging environment in order to investigate the sequential portions of the code relative to each process. For that reason we are trying to combine MAD with another debugger, e.g. [3]. This should lead to a powerful debugging environment, that supports users during most activities of parallel program debugging.

## Appendix: Summary of Prepared Demo

The interactive demonstration accompanying this paper will try to emphasize on the usefulness of our approach for parallel program debugging. For that reason, we provide two example programs implemented with the standard Message Passing Interface. One program is a simple parallel solver for Poisson's equation, the other is a code fragment of a distance doubling algorithm. The following steps are planned for the demo:

- Instrumentation of the parallel Poisson solver with NOPE (see Section 2).
- Execution of the program with MPI and generation of corresponding tracefiles.
- Visualization of the collected tracedata within ATEMPT (see Section 3).
- Demonstration of automatic error detection and array inspection (see Figure 1).
- Execution of the nondeterministic distance doubling program - different results will be observable, although the same input data is provided.
- Automatic generation of all possible program execution based on the initial execution (see Section 4).